\documentclass{article}
\usepackage{spconf,amsmath,graphicx}
\usepackage{xcolor}
\usepackage{booktabs}
\usepackage{tikz}
\usepackage{float}
\usepackage{hyperref}
\usepackage{fancyhdr}
\ninept
\title{REAL-TIME AUDIO-VISUAL END-TO-END SPEECH ENHANCEMENT}
\name{Zirun Zhu, Hemin Yang, Min Tang, Ziyi Yang, Sefik Emre Eskimez, Huaming Wang}
\address{Microsoft, Redmond, WA, USA\\{\small \texttt{\{zirzhu, heyang, mintang, ziyiyang, seeskime, huawang\}@microsoft.com}}}
\begin{document}
%
\maketitle
\begin{abstract}
Audio-visual speech enhancement (AV-SE) methods utilize auxiliary visual cues to enhance speakers' voices. Therefore, technically they should be able to outperform the audio-only speech enhancement (SE) methods. However, there are few works in the literature on an AV-SE system that can work in real time on a CPU. In this paper, we propose a low-latency real-time audio-visual end-to-end enhancement (AV-E3Net) model based on the recently proposed end-to-end enhancement network (E3Net). Our main contribution includes two aspects: 1) We employ a dense connection module to solve the performance degradation caused by the deep model structure. This module significantly improves the model's performance on the AV-SE task. 2) We propose a multi-stage gating-and-summation (GS) fusion module to merge audio and visual cues. Our results show that the proposed model provides better perceptual quality and intelligibility than the baseline E3net model with a negligible computational cost increase.
\end{abstract}
\begin{keywords}
speech enhancement, audio-visual, real-time, low-latency, dense connection
\end{keywords}
\section{Introduction}
\label{sec:intro}

Video communications have been universally applied for both business and personal connections due to the COVID-19 pandemic. Since numerous people have shifted to online communication, the request for better perceptual quality and intelligibility of audio has become increasingly crucial. However, some factors, such as background noise, reverberation, and interfering (background) speakers, can degrade the quality of the call. The leakage of interfering speakers can significantly degrade the intelligibility of the main speaker and potentially cause privacy issues. Unfortunately, unconditional audio-only speech enhancement models cannot remove interfering speakers since they are usually trained to preserve all human speech~\cite{c1,c3}. 


Personalized speech enhancement (PSE) models were proposed to enhance target speakers by adding target speakers' voice profiles~\cite{c1,c3,c2,c4} for suppressing the interfering speakers and environmental noises. They utilized the speaker embeddings extracted by a pre-trained speaker encoder on enrollment audios~\cite{c4}. Alternatively, video can assist in speech enhancement from different aspects without the need for enrollment. First, the face of the speaker reveals the speaker's identity~\cite{c5}. Second, lip motion is highly correlated with the phonetic information of the target speech~\cite{c6}.

\begin{figure*}[htb]

\begin{minipage}[b]{1.0\linewidth}
 \centering
 \centerline{\includegraphics[width=14cm, height=6.5cm]{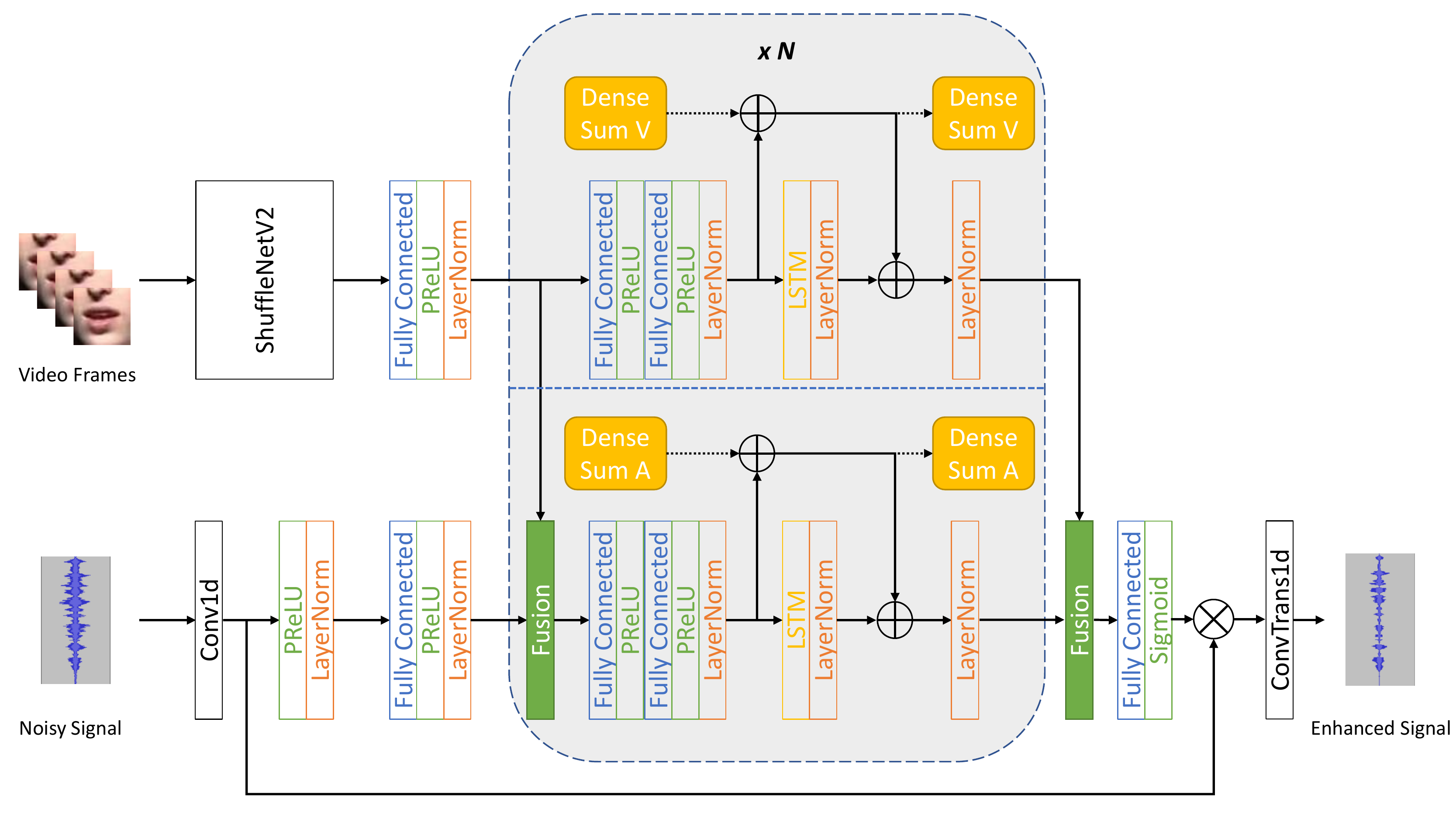}}
\vspace{-1.em}
\end{minipage}
\caption{Model architecture of proposed AV-E3Net. $\bigoplus$ denotes addition and $\bigotimes$ denotes Hadamard product. Dense sum A/V denotes the dense connection variable for audio/video features, which is depicted in subsection 3.1. The dense sum V provides shortcuts across all video LSTM blocks. The dense sum A provides shortcuts across all audio LSTM blocks as well as all fusion blocks. The fusion block can be either a concatenation block or a GS block.}
\label{fig:res}
\vspace{-1em}
\end{figure*}

It is essential to limit the latency and computational complexity of the model to make audio-visual speech enhancement (AV-SE) models suitable for real-time communication. Although many works employed causal designs, only a few reported the computational complexity of their methods~\cite{c7}. This paper focuses on optimizing intelligibility and perceptual quality for real-time processing on the CPU (i.e., the inference time on the CPU is shorter than the audio time with low latency). Specifically, we propose a low-latency real-time AV-SE system, namely AV-E3Net, based on the recently proposed end-to-end enhancement network (E3Net)~\cite{c1}. AV-E3Net takes pixels of the mouth region of interest (ROI) and the noisy audio signal as inputs and produces the enhanced audio signal. We employ a dense connection module in AV-E3Net, which helps with better gradient flow for deeper networks. We propose a novel multi-stage gating-and-summation (GS) fusion module for merging audio and video features. We evaluate our proposed models in different application scenarios. Our results suggest that AV-E3Net yields significantly better results for the AV-SE task than the baselines.    

\section{Related Work}

Recently, multiple personalized speech enhancement (PSE) systems were proposed and proven computationally efficient to work in real time. Personalized PercepNet~\cite{c2} utilized the target speaker's voice embeddings to improve the speech enhancement capabilities of original PercepNet \cite{valin20_interspeech}. Thakker et al.~\cite{c1} proposed a real-time causal PSE model named end-to-end enhancement network (E3Net). Compared with other bigger PSE models such as pDCCRN~\cite{c3}, E3Net provided better perceptual and transcription quality with much smaller computational complexity. PSE models are conditional models that utilize speaker embeddings, and their success can be extended to other conditional models, such as AV-SE models.

Traditionally, an AV-SE network comprises four components: audio encoder, video encoder, enhancement network, and audio decoder~\cite{c8,c9}. The audio/video encoder extracts audio/video features, and the enhancement network combines two features to produce enhanced audio embedding. The audio decoder decodes enhanced audio embedding to recover the audio. The video encoder can be pre-trained or jointly trained with the other modules. In~\cite{c8}, the video encoder was trained jointly with the rest of the network, whereas~\cite{c9} and~\cite{afouras18_interspeech} employed a pre-trained video encoder. Joint training of the video encoder with the rest of the network is somewhat challenging because of the deeper model architecture. However, there are some techniques to alleviate this problem. ResNet~\cite{c10} and DenseNet~\cite{c11} proposed dense shortcuts to address the training issue caused by the deep structures. Zhang et al.~\cite{c12} also proposed a unified perspective of the dense shortcut in ResNet and DenseNet. Motivated by these works, this paper employs a dense connection module to tackle the performance issue caused by the deep architecture of AV-E3Net. 

Audio and video fusion is an important research direction for AV-SE ~\cite{c8,c13,c15} and multimodal learning~\cite{c26}. The most common fusion method is concatenation, which is easy to implement, but one modality often tends to dominate the other~\cite{c8}. Xu et al.~\cite{c13} proposed an attention-based fusion method. However, this method considerably increased the computational cost. Joze et al.~\cite{c14}, and Iuzzolino and Koishida~\cite{c15} proposed a squeeze-excite (SE) fusion that employs a gating module to recalibrate the modality. Additionally, their work integrated slow fusion~\cite{c16} with the gating module and proved that slow fusion is more effective. Wang et al.\cite{9053033} also employed a gating network to perform a product-based fusion, ensuring the performance of the model lower-bounded by an audio-based system. Inspired by~\cite{c15,c14,9053033}, we propose a multi-stage gating-and-summation (GS) module, which integrates slow fusion and provides a lightweight and efficient approach to fuse audio and video features. 

An essential requirement for AV-SE systems to be adopted in practical scenarios is to make them work in real time with low computational costs. Unfortunately, only a few existing works focused on this scenario. Gu et al.~\cite{c7} proposed a real-time audio-visual speech separation and provided the real-time factor (RTF) measured on a GPU as the metric for computational costs. Gogate et al.\cite{GOGATE2020273} also proposed a real-time audio-visual speech enhancement model, whereas no computational complexity metric was provided. In contrast, we propose an AV-SE system that can work in real time on the CPU by utilizing computationally efficient E3Net as our backbone, using lightweight ShufflNetV2 as the video encoder, and using only mouth ROI as the visual input. 

\section{Methodology}
\label{sec:format}

The overview of AV-E3Net architecture is shown in Figure~\ref{fig:res}. In this section, we introduce each module of the proposed model.

\subsection{Audio network}
\label{ssec:audio nnet}
The audio network comprises an audio encoder, a masking network, and an audio decoder. The audio encoder processes the input audio to generate audio features; subsequently, they are fed into the masking network to produce a mask, which is applied to the audio features. Within the masking network, the audio features are fused with video features generated by the video encoder. At last, the audio decoder reconstructs the audio from suppressed audio features. We follow the design of E3Net~\cite{c1}. The audio encoder and audio decoders are 1D convolution and 1D transposed convolution layers, respectively. The masking network linearly stacks a ReLU activation, a layer normalization, a projection block, a fusion module, multiple LSTM blocks, and a mask prediction module. Please refer to~\cite{c1} for the detailed description of the LSTM block and the mask prediction. In addition, we employ the dense connection module to tackle performance issues caused by a deep model structure. A dense connection replaces the original skip connection of E3Net in each LSTM block. The dense connection also exists in the fusion module. Generally, a module with a skip connection can be expressed as:
\begin{equation}
Y_n=\theta(f_n(x_n)+x_n)
\end{equation}
where n is the index of the module, $\theta$ is a layer normalization, $x_n$ is the input of the  module $f_n$, and $Y_n$ is the output. As an alternative, the dense connection is defined as:
\begin{equation}X_n=\Sigma_{i=0}^{n}x_i,\end{equation}
\begin{equation}Y_n=\theta(f_n(x_n)+X_n)\end{equation}
or
\begin{equation}X_n=X_{n-1}+x_n,\end{equation}
\begin{equation}Y_n=\theta(f_n(x_n)+X_n)\end{equation}
Therefore, $X_n$ is a dense summation variable that is updated from $X_{n-1}$ to $X_n$ through all dense connection blocks. Note that a dense connection requires each $x_n$ to have the same shape. In this way, the dense connection provides shortcuts across all LSTM blocks and fusion modules.

\subsection{Video encoder}
\label{ssec:video encoder}
In a recent lipreading study, Ma et al.~\cite{c17} employed a lightweight ShuffleNetV2~\cite{c18} as the video encoder to extract visual features. This work verified that ShuffleNetV2 could provide satisfactory performance with much efficient computational complexity in lipreading tasks. Motivated by its success in lipreading, we also employ ShuffleNetV2, followed by a projection block, as the video encoder. The projection block changes the dimension of video features. It comprises a fully connected layer, a PReLU activation, and a layer normalization. Afterward, we optionally employ video LSTM blocks to capture the speaker's lip motions. Video LSTM blocks share the same structure as those in the audio network.

\subsection{Audio-Visual fusion}
\label{ssec:av fusion}

Next, we describe our proposed multi-stage gating-and-summation (GS) fusion module. For the multi-stage fusion, the video LSTM blocks are forced to be paired with the audio LSTM blocks in the masking network. The fusion blocks are placed at the beginning of each pair of LSTM blocks and after the last pair of LSTM blocks, as shown in Figure~\ref{fig:res}.

\begin{figure}[!t]

\begin{minipage}[b]{1.0\linewidth}
 \centering
 \centerline{\includegraphics[width=5cm]{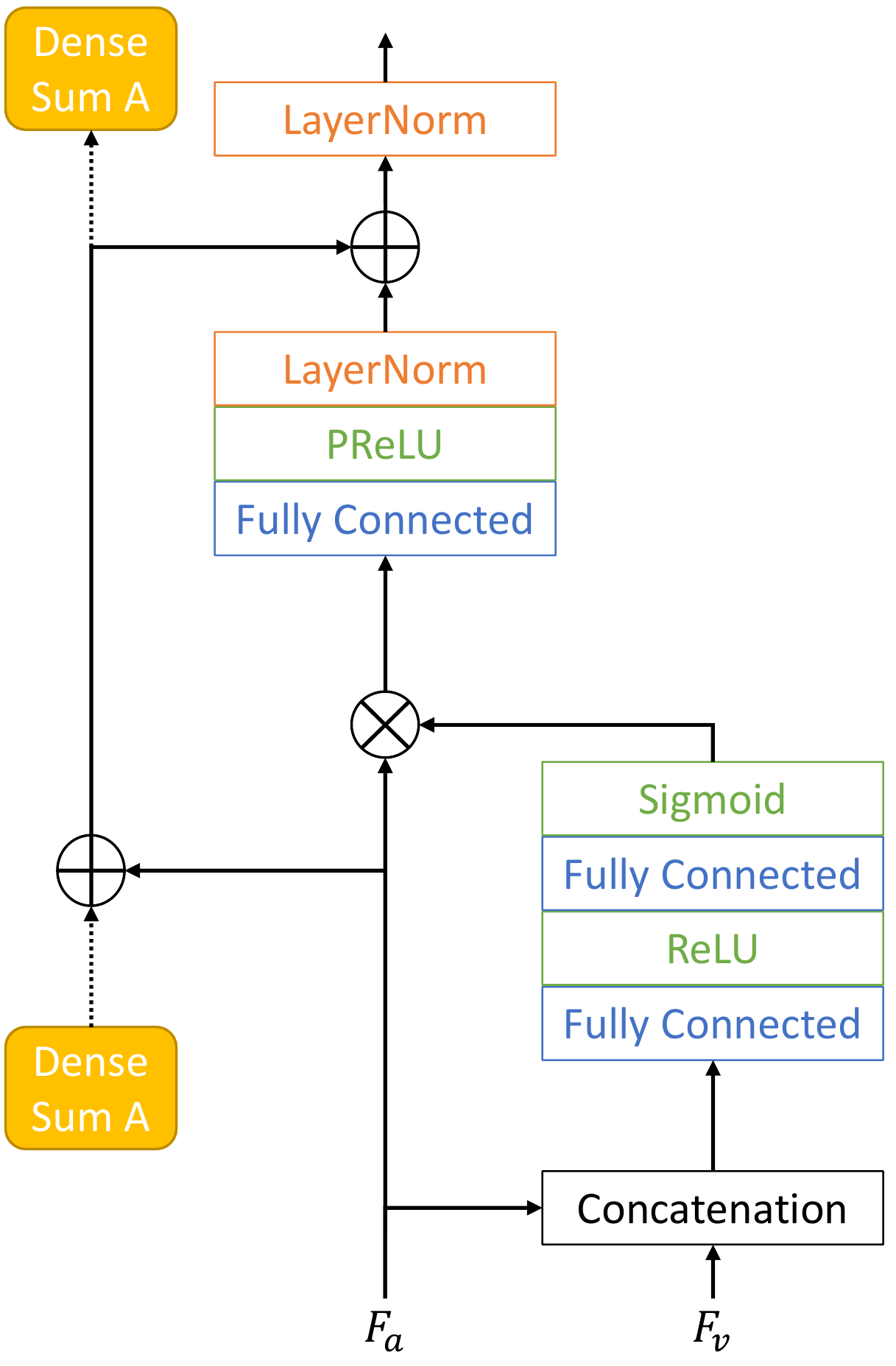}}
\vspace{-1em}
\end{minipage}
\caption{Proposed architecture of GS fusion block. $\bigoplus$ denotes addition and $\bigotimes$ denotes Hadamard product. Dense sum A denotes the dense connection variable for audio features.}
\label{fig:res1}
\vspace{-1em}
\end{figure}

Figure~\ref{fig:res1} presents the detailed architecture of the GS fusion block.
The block takes audio features $F_{a,n} \in R^{{d_a}}$ and video features $F_{v,n} \in R^{{d_v}}$ from the $n^{th}$ pair of LSTM blocks as inputs. Two operations are included in this fusion block. First, a gating module is employed to calculate the importance of channels and recalibrate the original audio features by the importance,
\begin{equation}G_n = \sigma(g(\delta(h([F_{a,n};F_{v,n}])))),\end{equation}
\begin{equation}H_{a,n} = G_n \odot F_{a,n}\end{equation}
where $[F_{a,n};F_{v,n}]$ concatenates audio features and video features, $G_n \in R^{d_a}$, $\sigma$ is the sigmoid activation, h and g are fully connected layers, and $\delta$ is a ReLU activation. Second, a dense summation module is defined as,
\begin{equation}X_{a,n}=X_n+F_{a,n}\end{equation}
\begin{equation}Y_{a,n} = \theta(proj(H_{a,n})+X_{a,n})\end{equation}
where $proj$ is a projection block, $\theta$ is the layer normalization, $X_n$ is the dense summation variable from the $n^{th}$ audio LSTM block. Updated by adding $F_{a,n}$, the new dense summation variable becomes $X_{a,n}$.~\cite{9053033} observed serious performance degradation of other AV-SE models~\cite{c8,c9} in less noisy scenarios. Therefore, they suggested that AV-SE systems should be mainly audio-based, and video cues should provide only additional contributions. The design of GS fusion follows this idea: the gating module and the summation module make fused features closer to the space of audio features. The dense summation also provides shortcuts across all LSTM blocks and fusion modules and helps with better gradient flow for AV-E3Net.

Note that, due to the mismatch of frame rates for the video and audio, we up-sample the video frames to match them with the audio frames by replicating them. 


\section{Experimental Results}
\label{sec:pagestyle}

\subsection{Training and validation data}
\label{ssec:data}
This work followed the data simulation pipeline of~\cite{c3}. We utilized clean speech samples from AVSpeech~\cite{c9}, VoxCeleb2~\cite{c19}, and LRS3~\cite{c20} datasets. By filtering the samples according to the Mean Opinion Score (MOS) of the audio quality, we selected 214, 170, and 30 hours of video samples from these datasets, respectively. Furthermore, we used a face detector for each frame of the video samples, and if the number of frames with a face detected divided by the total number of frames was less than 0.95, we discarded that video sample. 

For creating the noisy mixtures, we convolved clean speech samples with simulated room impulse response (RIR) using the image method. We employed noise clips from Audioset~\cite{c21} and Freesound~\cite{c22}, which were also convolved with RIRs. The clean speech, noise clips, and RIR files were split exclusively to simulate train, validation, and test sets. 20$\%$ of simulated samples contained only the target speaker, and 80$\%$ of them contained both the target and interference speakers. In~\cite{c3}, there was a restriction that the target speaker should be closer to the microphone than the interference speaker. However, in this work, we relieved this restriction. Our system assumes that the target speaker's face is the only face captured by the camera. We simulated 20,000 hours of training data and 10 hours of validation data. The average length of simulated mixture samples was around 10 seconds. We used the video frames unaltered along with simulated noisy audio samples. For video frames where the face detector did not capture the target speaker's face, we filled in the video frame input with zero tensors. 

\begin{table*}[htb]
\caption{Computational complexity and model performance results are as shown. RTF was measured on an Intel(R) Xeon(R) W-2133 CPU @ 3.60GHz and averaged on 100 runs. The multi-stage fusion is introduced in subsection 3.3. "single concat" denotes the single concatenation block which is introduced in 4.4.}
\vspace{.2em}
\centering
\resizebox{\textwidth}{!}{
\begin{tabular}{lccccccccccccc} \hline
\multicolumn{3}{c}{\textbf{Configuration}}&& \multicolumn{2}{c}{\textbf{Complexity}}&& \multicolumn{3}{c}{\textbf{TS1}} &&  \multicolumn{3}{c}{\textbf{TS2}}                                                                  \\ \cline{1-3}\cline{5-6} \cline{8-10} \cline{12-14} \cline{13-14}
Method & Dense connection & Fusion & & Parameters(millions) & RTF  & & WER $\downarrow$        & SDR   $\uparrow$       & PESQ $\uparrow$           & & WER  $\downarrow$      & SDR $\uparrow$           & PESQ  $\uparrow$     \\\hline
No Enhancement &  & &  &  &  & &24.38 & 3.26 & 1.186 & & 14.43 & 6.53 &1.294 \\ \hline
AO-E3Net &no & no  &  & 16.03 &0.053& & 26.34 & 6.70 & 1.451 &  & 17.32 & 12.05 & 1.930 \\
AV-DCATTUNET & no & single concat &  & 10.87 &1.580& & 17.93 & 10.72 & 1.946 &  & 14.79 & 12.41 & \underline{2.158} \\ 
Naive AV-E3Net &no & single concat & & 18.02 & 0.122 & & 18.11 & 11.41 & 1.958 & & 15.51& 12.67 & 2.082 \\ 
\hspace{0.4cm} - w/ 1 video LSTM block & no & single concat &  &21.17 & 0.127 &  & 18.06 & 11.38 & 1.958 &  & 15.24& 12.70 & 2.096 \\
\hspace{0.4cm} - w/ 4 video LSTM blocks & no & single concat &  &30.64 & 0.138 &  & 24.86 & 10.20 & 1.806 &  & 18.58& 12.63 & 2.061 \\\hline
AV-E3Net &yes & single concat & & 18.02 & 0.123 & & 17.01 & 11.52 & 1.974 & & 14.76& 12.72 & 2.099 \\ 
\hspace{0.4cm} - w/ 1 video LSTM block & yes & single concat &  &21.17 & 0.130 &  & 16.95 & 11.54 & 2.000 &  & 14.87& 12.76 & 2.136 \\
\hspace{0.4cm} - w/ 4 video LSTM blocks & yes & single concat &  &30.64 & 0.138 &  & 16.73 & 11.57 & 1.985 &  & 14.35& 12.73 & 2.106 \\\hline
AV-E3Net w/ 4 video LSTM blocks &yes & multi-stage & & 32.74 & 0.142 & & 16.64 & 11.60 & 2.002 & & 14.19& 12.78 & 2.104 \\ 
\hspace{0.4cm} - GS fusion (proposed) &yes &multi-stage&   &35.37 & 0.143 &  & \textbf{16.62} & \textbf{11.67} & \textbf{2.009} &  & \textbf{14.02}& \textbf{12.83} & \textbf{2.136} \\\hline
\end{tabular}
}
\label{tab:results}
\vspace{-0.5cm}
\end{table*}

\subsection{Test sets and evaluation metrics}
\label{ssec:test}
Test sets followed the same simulation approach as train/validation sets, in which the source data were mutually exclusive. Only LRS3~\cite{c20} data was used in test sets, and the average length of simulated mixture samples was 6 seconds. We simulated test sets for two target scenarios: 1) the mixture comprises the target speaker, the interference speaker, and noise. 2) the mixture comprises the target speaker and noise. Following~\cite{c3}, we named these two scenarios TS1 and TS2, respectively. TS1  and TS2 included 10 hours and 1 hour of data, respectively. To measure the perceptual quality and the intelligibility of processed audio, we utilized word error rate (WER), perceptual evaluation of speech quality (PESQ), and Signal-to-distortion ratio (SDR). We also measured the real-time factor (RTF) on an Intel(R) Xeon(R) W-2133 CPU @ 3.60GHz. Since the inference speed of the model changed from time to time on a CPU, we ran the same model 100 times on a 3 seconds input to reduce the variance of observations.

\subsection{Implementation Details}
\label{ssec:settings}
Audio and video samples were re-sampled in 16KHz, 25 fps, and 360p, respectively. We followed video pre-processing of lipreading~\cite{c17,c23}. Each video frame was processed by 1) face and landmarks detection, 2) similarity transformation based on landmarks, and 3) cropping in a size of 50x50 on the mouth ROI. Small sizes for cropping can further reduce the video encoder's computational cost, which contributed most of the computational cost in AV-E3Net. We used Microsoft's internal face detection tool for face and landmarks detection. During training, the noisy mixtures were chunked into 3 seconds of audio batches aligned with 75 video frames. We used the power-law compressed phase-aware (PLCPA) loss function~\cite{c24}.

Regarding the audio encoder and decoder, we set window and hop sizes to 320 (20 ms) and 160 (10 ms), respectively. The theoretical latency of AV-E3Net was 20ms. The number of features used in the audio encoder was 2048. Within the masking network, the projection block projected features from $R^{2048}$ to $R^{512}$. The number of audio LSTM blocks was 4. Within the LSTM block, the input and output dimensions of the fully connected block were 512, and the intermediate dimension of the fully connected block was 1024. The input and output dimensions of LSTM were 512. Regarding the video encoder, we used ShuffleNetV2 0.5x~\cite{c18} to encode video frames to 1024-dimension features. Then a projection block was employed to project video features to 512-dimension. Afterward, video LSTM blocks for lip motion capture shared the same configuration as audio LSTM blocks. GS fusion's audio and video input dimensions were 512, and the first fully connected layer projected concatenated features from $R^{1024}$ to $R^{512}$. We set the optimizer as AdamW~\cite{c25} and the learning rate scheduler to be a step decay scheduler with a gradual warm-up mechanism. The peak learning rate was 0.001.

\subsection{Baseline Systems}
We employed the following baseline models for comparison with our proposed models: 

\noindent\textbf{AO-E3Net:} An audio-only E3Net model. It is an unconditional model and not capable of removing the interfering speaker.

\noindent\textbf{AV-DCATTUNET}: A variant of pDCATTUNET, introduced in~\cite{c3}, in which the speaker embedding was replaced by the video frame embedding extracted by a pre-trained face recognition model (ShuffleNetV2 0.5x). The face recognition model takes the whole face of each video frame as the input. We set the number of encoder/decoder blocks to 6 and the number of bottleneck blocks to 4. The STFT window and hop sizes were 512 and 256 samples, respectively.

\noindent\textbf{Naive AV-E3Net:} The AV-E3Net without dense connection and multi-stage GS fusion. It combined the video encoder (ShuffleNetV2 0.5x) with E3Net and employed a single concatenation block to merge late video features from the video encoder with intermediate audio features before the first audio LSTM block. A single concatenation block comprises a concatenation layer,  a projection block, a dense connection or a skip connection, and a layer normalization. Particularly, only skip connection was used in Naive AV-E3Net's LSTM block and single concatenation block.

\subsection{Results}
\label{ssec:results}

Table~\ref{tab:results} shows the computational complexity of different model configurations and the corresponding perceptual quality and intelligibility results on TS1 and TS2 test sets. According to the results, AO-E3Net performs poorly on TS1 since it cannot remove the interfering speaker. In contrast, AV-DCATTUNET provides much better results on both TS1 and TS2 than the AO-E3Net in terms of speech and transcription quality. Naive AV-E3Net without video LSTM yields worse WER results than AV-DCATTUNET but provides better SDR. Adding a single video LSTM to the Naive AV-E3Net yields similar results; However, increasing it to 4 LSTM blocks degrades speech and transcription quality, indicating training difficulty. By adding the dense connection to AV-E3Net, we observe significant speech and transcription quality improvement with a negligible computational cost increase. With the dense connection, adding more video LSTM layers improves the transcription quality while yielding similar speech quality. AV-E3Net models with dense connection outperform AV-DCATTUNET on TS1 and achieve comparable performance on TS2 with a much lower computational cost. Next, the results show that AV-E3Net with multi-stage training further improves speech and transcription quality. The best AV-E3Net results are obtained using multi-stage fusion with the GS fusion. The GS fusion helps with substantially better performance on TS2, which is for the less noisy scenario. The computational cost increase for multi-stage GS fusion is minor.

It should be noted that the AV-DCATTUNET model cannot be used for real-time processing because of the model's dependency on a pre-trained video encoder. Although the face recognition model employs the efficient ShuffleNetV2, it takes the whole face rather than the mouth ROI as the input. Therefore, the RTF on the video encoder reaches 1.442. However, since AV-E3Net uses only mouth ROI as the input, it can work in real time. These results suggest that AV-E3Net performs better than the bigger model (AV-DCATTUNET) with a lower computational cost. 

\section{Conclusions}
\label{sec:typestyle}

We proposed a low-latency real-time audio-visual end-to-end speech enhancement model AV-E3Net. We employed a dense connection module, which significantly improved both perceptual quality and intelligibility with minimal increase in computational costs. Furthermore, we proposed a novel multi-stage gating-and-summation (GS) fusion module that dynamically and effectively fuses speech and vision modalities. We showed that our proposed model performed much better than the baseline systems. Furthermore, the ablation study showed the impact of adding dense connection and multi-stage GS modules. The computational cost of our system is much lower than the baseline AV-SE system and can work in real time on the CPU. Therefore, the proposed AV-E3Net has excellent potential in real-world video communication applications as a low-latency and real-time model.\footnote{Samples available at \url{https://github.com/zzrdwj/AVSE}}


\vfill\pagebreak

\bibliographystyle{IEEEtran}
\bibliography{refs}

\end{document}